\documentclass{tPHM2e}
\usepackage{epstopdf}
\begin{document}

\title{Static and dynamic glass-glass transitions: a
mean-field study} \author{Luca Leuzzi\thanks{Email:
luca.leuzzi@cnr.it, www.smc.infm.it}\\ \vspace{6pt} SMC Center,
INFM-CNR and Dept. of Physics, University ``Sapienza'' of Rome,\\ P.le
A. Moro 2, 00185, Rome, Italy}

\maketitle

\begin{abstract}
The behavior of a family of mean-field glass models is reviewed. The
models are analyzed by means of a Langevin-based approach to the
dynamics and a Replica theory computation of the thermodynamics.  We
focus on the phase diagram of a particular model case, where
glass-to-glass transitions occur between phases with a different
number of characteristic time-scales for the relaxation processes. The
appearence of Johari-Goldstein processes as collective reorganizations
of sets of fast processes is discussed.
\end{abstract}
\section*{}
Theoretical modeling of glassy systems is a widespread
topic. Different important theories have been introduced along the
years like, e.g., the ``free volume'' \cite{Turnbull61}, the
``entropic'' \cite{Gibbs58} and the ``random first order''
\cite{Lubchenko07} theories, to mention a few. Nevertheless, a
comprehesive theory both being analytically treatable and yielding
reliable quantitative predictions above and below the glass transition has
yet to be devised and appears to be a very complicated and challenging
aim.  In front of such limitations a theoretical approach based on the
mean-field approximation, where statistical fluctuations of
microscopic observables are neglected, helps pointing out a way to
enforce realistic approaches and indentifying physically relevant
concepts.

We present the study of a family of mean-field models with
time-independent, i.e., {\em quenched}, disorder.  The typical feature
of amorphous systems is the impossibility to reach states at the
lowest feasible energy, thus preventing crystallization.  The system,
undergoes some kind of ``frustration'': because of dynamic
arrest the global set of the energetic contributions due to the
interactions among the glass former constituents cannot be
simultaneously minimized.  In the present case the frustration is a
direct consequence of the quenched disorder. More generally it is
self-generated by the inner geometry of the material and/or by the
complicated exchange of interactions. The quenched disorder is not a
necessary ingredient,\footnote{Glassy models
without quenched disorder can be devised as well, see, e.g.,
Ref. \cite{Marinari94}.}  then, even though it makes the problem more
easily tractable.
The motivations for this study are manifolds.  First, applying the
replica method and using the concept of Replica Symmetry Breaking
(RSB), the analysis of thermodynamic and dynamic properties can be
carried out analytically.  Then, it is possible to develop and check a
multi-timescales equilibrium dynamics consistent at all times,
including the asymptotic limit.  Further, the model displays a very
rich phase diagram with different glass and spin-glass phases,
allowing for a theoretical analysis of the phenomenon of
polyamorphism.\footnote{Many examples of polyamorphism are available
in nature (and in literature).  For example, the change in the
kinetics of the coordination between molecules, occurring in vitreous
Germania and Silica \cite{Tsiok98} or the sharp density change taking
place in porous silicon \cite{Deb01}, as well as in undercooled water
\cite{Poole92}.  Very recently polyamorphism in Ethanol
\cite{Ramos08}, Laponite \cite{Ruzicka08} and star polymer mixtures
\cite{Mayer08} has been observed.}  Eventually, the identification of
 RSB's in the thermodynamics with time-scale separations in the
dynamics \cite{Sompolinsky81} provides an useful theoretical tool to
study the interrelation between primary ($\alpha$), secondary (or
Johari-Goldstein, $\beta_{\rm JG}$) and tertiary processes
($\beta_{\rm fast},\gamma$).  In this paper we will deepen the last
two aspects and their possible implications for real structural glasses.
%

We first briefly introduce the model and sketch the computation of its
thermodynamics within the Replica theory, emphasizing the nature of
the order parameter and its change in behavior accross qualitatively
different amorphous phases.  The model Hamiltonian is:
\begin{equation}
\label{f:Ham}
{\cal H} = \! \sum_{i_1<\ldots <i_s}\!J^{(s)}_{i_1\ldots i_{s}}
\sigma_{i_1}\cdots\sigma_{i_s}
           +\!\sum_{i_1<\ldots <i_p}\!J^{(p)}_{i_1\ldots i_p}
           \sigma_{i_1}\cdots\sigma_{i_p}
\end{equation}
where $ J^{(t)}_{i_1\ldots i_{t}}$ ($t=s,p$) are uncorrelated, zero mean,
Gaussian variables of variance
$J_t^2 t!/(2N^{t-1})$
and $\sigma_i$ are $N$ ``spherical spins'' obeying the 
constraint $\sum_i \sigma_i^2 = N$.

In a complex Free Energy Landscape (FEL), such as the one representing
an amorphous system, the numerous valleys, i.e., the ``glass states'',
can be more or less correlated among them and a hierarchy can be
established based of their relative correlation.\footnote{A well known
example is the symbolic dynamics through the Potential Energy
Landascape, where intra-basin processes have a high correlation and
inter-basin processes have a low correlation
\cite{Sciortino05,Leuzzi07}.}  Denoting by $\langle \ldots\rangle_a$
the thermal average over the configurations belonging to state
``$a$'', the following {\em overlap} order parameter is defined as the
correlation between two states ($a$ and $b$):
\begin{equation}
  q_{ab}=\frac{1}{N}\sum_{i=1}^N\langle\sigma_i\rangle_a 
\langle \sigma_i\rangle_b
\label{f:q}
\end{equation}
 To be precise, the complete order parameter is the {\em probability
distribution} $P(q)$ of the values of $q$ \cite{MPV87}. Depending on
the shape of $P(q)$ one can identify a specific phase of the amorphous
system. In table \ref{tab:phases} we summarize the most common
behaviors known in literature.
\begin{table}
\tbl{Phases and order parameters for models with quenched
disorder.}
{\begin{tabular}{|l|c|}
PHASE           & OVERLAP DISTRIBUTION \\
\hline
\vspace*{-1mm} &\vspace*{-1mm} \phantom{a}
\\
Paramagnet/Fluid \hspace*{1.2cm}& $ P(q)=\delta(q) $ \\
Glass            \hspace*{2cm}   & $P(q)= m \delta(q-q_0)+ (1-m)\delta(q-q_1)$ \\
Spin-Glass$^\ast$       \hspace*{2cm}   &  \hspace*{1.5cm}  $P(q)= w_0 \delta(q-q_0)+ \tilde P(q) + w_1\delta(q-q_1)$  \hspace*{1.5cm}
\\ \hline
\end{tabular}}
\tabnote{$^\ast$ The function $\tilde P$ is continuous on
the support $]q_0:q_1[$, $w_0$ and $w_1$ are weights of the
$\delta$'s ($w_0+w_1<1$).}
\label{tab:phases}
\end{table}

\begin{figure}[t!]
\begin{center}
  \includegraphics[width=.7\textwidth]{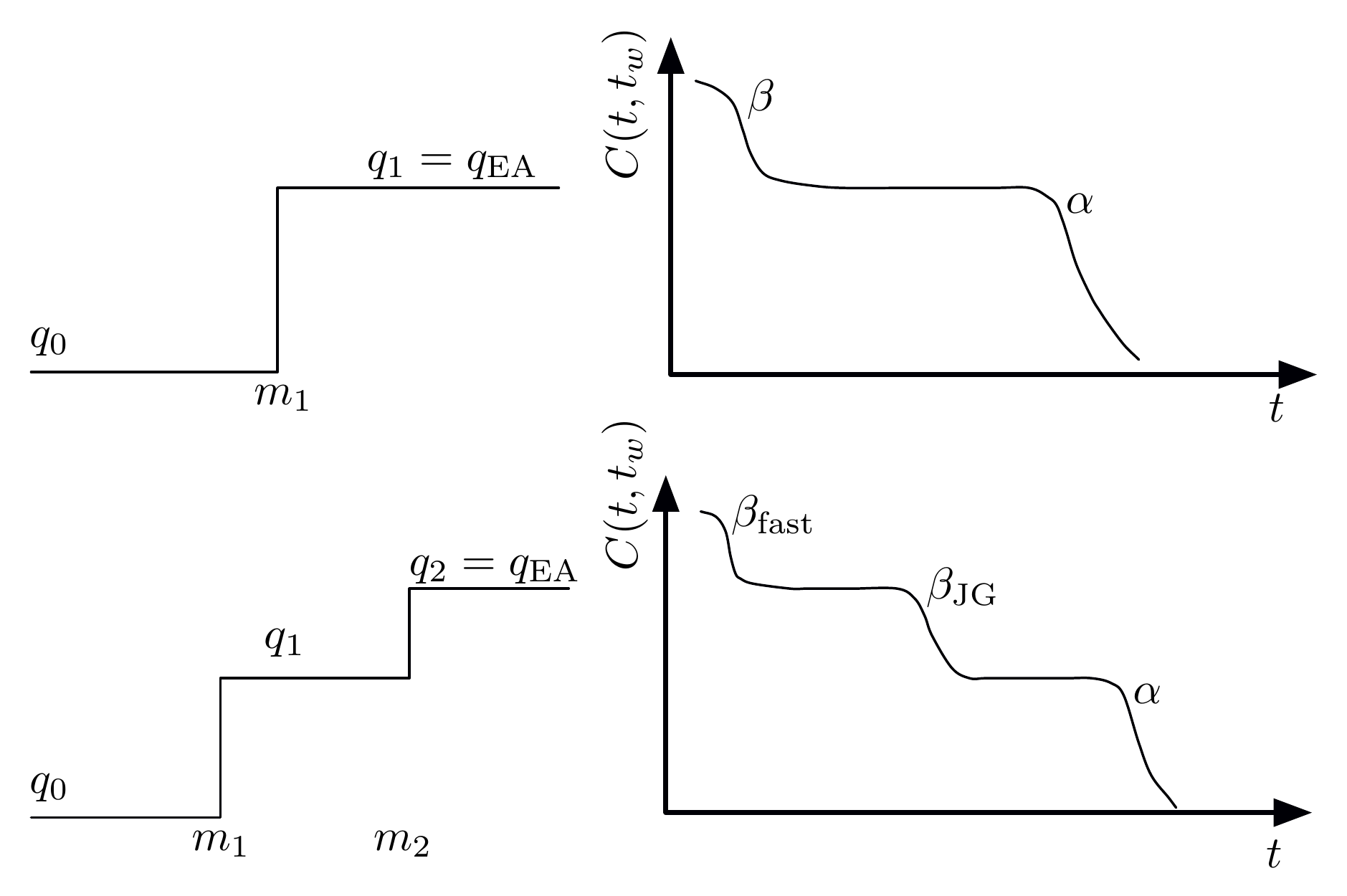}
\end{center}
\vskip  -3 mm
 \caption{L.h.s.: order parameter step function $q(x)$
for the 1 and 2 RSB thermodynamic glassy phases. R.h.s.:
Correlator vs. time for systems with processes relaxing on one and two
well separated time-scales.  A correspondence between the properties
of the overlap function in the static solution and the behavior of the
correlation function in the dynamics is shown (see also
Eq. (\ref{f:plat})).}
\label{fig:stat_dyn}
\end{figure}

The replica theory for mean-field disordered systems is applied
to compute the free energy functional
\cite{Crisanti04,Crisanti08}:
\begin{eqnarray}
-\beta \Phi&=& \frac{1}{2}(1+\ln 2\pi) +\frac{1}{2}\lim_{n\to 0}\frac{1}{n}
\sum_{ab}^{1,n}g(q_{ab})+\ln\det {\hat{\bm q}}
\label{f:Phi}
\end{eqnarray}
where ${\hat{ \bm q}}=\{q_{ab}\}$ is the Parisi overlap matrix. The
model is specified by the function $g(q) \equiv q^s\mu_s/s +
q^p\mu_p/p$, with $\mu_s=s \beta^2J_s^2/2$. For a generic RSB Ansatz
with $R$ breakings the elements of the Parisi matrix take values
$0=q_0<q_1<\ldots<q_R<q_{R+1}=1$ with relative multiplicities
$n=m_0>m_1>\ldots> m_R>m_{R+1}=1$. As, in the Replica computation,
$n\to 0$, the parameters $m_r$ acquire real values ($\in [0,1]$)
\cite{MPV87} and one can express the set of $q$ and $m$ values as a
(step) function $q(x)$.  Here we are interested in structural glass.
We will, thus, take into account model cases displaying phases with
one and two step RSB, whose overlap functions are schematically
represented on the left hand side of Fig. \ref{fig:stat_dyn}. These
can be qualitatively connected with real glass formers in which only
primary ($R=1$) or also secondary ($R=2$) processes are present.
\footnote{From the point of view of Replica
calculation we stress that a thermodynamically consistent example of a
2RSB phase has not been realized in models other than the $s+p$
spherical models \cite{Crisanti08}.}
Such glass models are realized taking 
$s>2$ and large $p-s$ \cite{Krak07,Crisanti07}.
Eq. (\ref{f:Phi}) for a $R=2$ RSB phase can be written as
\begin{eqnarray}
  -2 \beta \Phi&=&1+\ln 2\pi +
           g(1)+m_2[g(q_2)-g(q_1)]+m_1[g(q_1)-g(q_0)]
              \\
&+& \ln\chi_2-\frac{1}{m_2}\ln \frac{\chi_2}{\chi_1}-\frac{1}{m_1}
	    \ln\frac{\chi_1}{\chi_0}+\frac{q_0}{\chi_0}
\end{eqnarray}
with $\chi_2=1-q_2$, $\chi_1=\chi_2+m_2(q_2-q_1)$ and
$\chi_0=\chi_1+m_1(q_1-q_0)$.

In the $s+p$ models there are also different solutions (depending on
the values of $s$ and $p$ and of $T$, $J_s$ and $J_p$), 
displaying both continuous and discontinuous (and mixed) overlap functions.
\cite{Crisanti04, Crisanti08}.\footnote{Usually, a glass phase is
associated with discontinuous steps in the overlap, corresponding to a
sharp separation of time-scales.  A spin-glass phase is, instead,
characterized by a fully continuous $q(x)$.}

The order parameter function (i.e., the set of values of $m$'s and $q$'s) is
obtained by solving the following set of self-consistency equations:
\begin{eqnarray}
\label{f:self_g}
g(q_2)-g(q_1)&=&(q_2-q_1)\Bigl[
\Lambda(q_1)-\frac{1}{m_2\chi_1}
\Bigr]
-\frac{1}{m_2^2}\ln\frac{\chi_2}{\chi_1}
\\
g(q_1)-g(q_0)&=&(q_1-q_0)\Bigl[
\Lambda(q_0)-\frac{1}{m_1\chi_0}
\Bigr]
-\frac{1}{m_1^2}\ln\frac{\chi_1}{\chi_0}
\\
\Lambda(q_0)=\frac{q_0}{\chi_0^2};&&\quad
\Lambda(q_1)-\Lambda(q_0)=\frac{q_1-q_0}{\chi_0\chi_1};
\qquad
\Lambda(q_2)-\Lambda(q_1)=\frac{q_2-q_1}{\chi_1\chi_2}
\label{f:self_q}
\end{eqnarray}
with $\Lambda(q) = d g(q)/dq$.
The thermodynamics of the 1RSB solution is obtained from the above
equations setting $q_2=q_1$.

The value of the overlap corresponds to a given correlation among states,
cf. Eq. (\ref{f:q}). The three levels function displayed in the 2RSB solution
corresponds to a precise hierarchy in the organization of the
thermodynamically relevant glassy states, consisting in groups of
states (clusters) and groups of gropus of states (``meta''-clusters).  Two
states whose overlap is $q_2$ belong to the same cluster.  Two states
whose overlap is $q_1$ do not belong to the same cluster but to the
same meta-cluster. Eventually, two states whose overlap is $q_0$
(usually equal to zero in absence of external forces of fields) belong
to different meat-clusters.


The dynamics of the model is Langevin.  Using a Martin-Siggia-Rose
 path-integral formalism  one can reduce the equations of motion
to a single variable formulation. Details can be found in
Ref. \cite{Crisanti07}.  The most important two-time observables are
the correlation and response function
\begin{equation}
C(t,t_w)={\overline{\langle\sigma(t)\sigma(t')\rangle}}; \qquad \qquad
G(t,t_w)=\frac{\delta {\overline{\langle \sigma(t)\rangle}}}{\delta
\beta h(t')} \quad t\geq t_w
\end{equation}
Let us take an amorphous phase with a generic number $R$
($=1,2,\ldots$) of time-scale bifurcations and assume that equilibrium
is obtained in each (completely) disjoint time-sector, i.e.  $t_w\to
-\infty$.  Using, e.g., a multiple scale analysis one assumes that
 the correlation function $C(t)$, as well as the response
$G(t)$, can be represented as the sum of $R+1$ distinct terms each
depending on a time variable, $\tau_0\ll \ldots \tau_r
\ll\ldots\ll\tau_R$, describing the motion in a given time-sector:
\begin{equation}
C(t) = \sum_{r=0}^{R}\,C_r(\tau_r),\quad 
\end{equation}
Considering a time-sector $a$ means to probe the dynamics on times
$t\sim \tau_a$.  We can split off the $a$-sector function $C_a$
 taking the ordered limit 
\begin{equation}
\widehat{\lim}_{t\to\infty}\equiv
\lim_{\tau_{R} \to \infty} \ldots \lim_{\tau_0\to\infty}
,
\end{equation} 
with the prescription $\tau_a/t=O(1)$, $\tau_{r<a}/t\to 0$ and
$\tau_{r>a}/t\to\infty$.  In practice, all contributions $C_r$ to the
correlation function with $r<a$ correspond to processes already
thermalized at the observation time $t$, whereas all contributions
whose index is larger than $a$ represent processes that are frozen at
time $t$.  The interesting processes under probe are those {\em
relaxing on characteristic times} $\tau_a\sim t$.  In the above
formulation, the asymptotic value of the correlator is, then, $C(t)\to
q_r$, and we have the condition:
\begin{equation}
\widehat{\lim}_{t\to\infty}\Biggl[~~\sum_{s=0}^{r-1}\,C_s(\tau_s) +
       \sum_{s=r}^{R}\,C_s(\tau_s)~~\Biggr] = q_r \quad \qquad \forall
       r=0,\ldots,R
\label{f:plat}
\end{equation}
A schematic behavior of $C(t)$ for the cases of our interest, $R=1,2$,
 is plotted on the r.h.s. of Fig. \ref{fig:stat_dyn}, next to their
 overlap counterparts. In the top part we have the thermodynamic order
 parameter (step) function $q(x)$ displaying a single discontinuity at
 $x=m_1$.  The two segments of the step function ($q_0,q_1$) can be
 linked, in the dynamics, to the two plateaus of the relaxation
 function in ordinary glass formers, in cases where secondary
 processes play no role.  In the bottom part the 2RSB case is
 sketched, i.e., two discontinuities in $q(x)$ (thermodynamics), or
 two time-scale separations in $C$ (dynamics). This is likely to be
 the mean-field reduction of a glass with secondary processes.

The response function on multiple separated time-scales reads
\begin{equation}
\label{eq:grtCHS}
G(t) = \sum_{r=0}^{R}\, \frac{\tau_r}{t}\,G_r(\tau_r)
\end{equation}
where each function $G_r$ varies only in the corresponding sector $r$,
 $\tau_r\sim O(t)$ and vanishes in all sectors with $s<r$.  The
 function $G_r$ represents the response of the system to a
 perturbation in the time sector labeled by $r$, i.e., the response
 due to all degrees of freedom which have not equilibrated in previous
 sectors.

Working with the Fourier transforms of the correlation and response
functions and defining the kinetic coefficient
$\Gamma^{-1}(\omega)=i\partial_\omega G^{-1}(\omega)$, the dynamical
stability is guaranteed by the requirements $\Gamma(\omega_r)=0$,
$\forall r=1,\ldots, R$ and $\Gamma(\omega_0)>0$, as the ordered limit
$\lim_{\omega_R\to 0}\ldots\lim_{\omega_0\to 0}$ is performed.
In our $R=2$ case the   conditions can be written as
\begin{eqnarray}
\Lambda'(q_2)=1/\chi_2^2, \qquad  
\Lambda'(q_1)=1/\chi_1^2, \qquad 
\label{f:dyn_stab}
\qquad
\Lambda'(q_0)>1/\chi_0^2 
\end{eqnarray} 
and provide the equations for the asymptotic dynamic solution. We
stress that the solution to Eq. (\ref{f:dyn_stab}) do not coincides
with the static solution, Eqs. (\ref{f:self_g})-(\ref{f:self_q}). This
is {\em typical of systems undergoing a dynamic arrest} before
reaching a temperature where they can undergo a thermodynamic phase
transition.  The thermodynamic transition in these spin-glass inspired
mean-field models for the glass is, instead, the Kauzmann transition
(at $T_K$), whereas the dynamic transition (at $T_d$) is equivalent to
the dynamic arrest transition predicted, e.g., in Mode Coupling Theory
(MCT). In real experiments it corresponds to the crossover temperature
at which the separation of time-scales of slow and fast processes
accelerates. \footnote{We notice that the experimental,
calorimetric, glass temperature $T_g$ is not defined in mean-field
systems. Indeed, this is a property connected with the falling out of
equilibrium of activated processes (hopping among valleys), whereas in
mean-field metastable states are surrounded by infinite barriers (as
$N\to\infty$). $T_g$ lies, undetermined, between $T_K$ and $T_d$.}

 A straightforward link with schematic models in MCT \cite{Fuchs91}
can be drawn, starting from the observation that the dynamic equations
in random spherical models are equivalent to the MCT equations
\cite{Bouchaud96} at high temperature, where time translational
invariance (TTI) holds and $G$ and $C$ are connected by the
fluctuation-dissipation theorem (FDT),
$G(t-t')=-\beta\theta(t-t')\partial_t C(t-t')$.  In our model case,
thus, if we take a memory kernel depending on the correlator $\phi$
(in MCT notation) as
$m(\phi)=\mu_s \phi^{s-1}+\mu_p\phi^{p-1}=\Lambda(\phi)$,
the mode coupling equations describe the Langevin dynamics of the
$s+p$ model and the overlap is identified with the non-ergodicity
parameter: $q=\lim_{t\to\infty} \phi(t)$.  The two dynamics differ,
instead, below $T_d$, since the global TTI breaking is implicit in the
random model dynamics, and FDT does not apply anymore in the above
form above.\footnote{For details on the generalization of
equilibrium dynamics in the solid amorphous phase see
Ref. \cite{Crisanti07}).}


 In mean-field models, unlike real glasses, the configurational
entropy $S_c={\overline{\log {\cal N}_J}}$ is a true state function.
 It can be formally computed as the
Legendre transform of the total free energy $\Phi$, with $f$ and
$\beta m$ as conjugated variables:
\begin{equation}
S_c(f;T)/N=\min_{m}\Bigl[-\beta m \Phi(m;T)-\beta m f\Bigr]
\end{equation}
where $m=m_R$. 
  The configurational entropy of the thermodynamic solution is
subsextensive ($S_c(f_{\rm eq})/N\to 0, N\to \infty$, Kauzmann point).
By maximizing $S_c$ vs. $f$, instead, we find the same values of $q_r$
and $m_r$ that solve the dynamic equations.  As a consequence, {\em
the dynamic arrest temperature can also be identified by looking at
the temperature at which an extensive configurational entropy arise}.

\begin{figure}[t!]
\begin{minipage}[b]{.52\textwidth}
\begin{center}
\includegraphics[width=\textwidth]{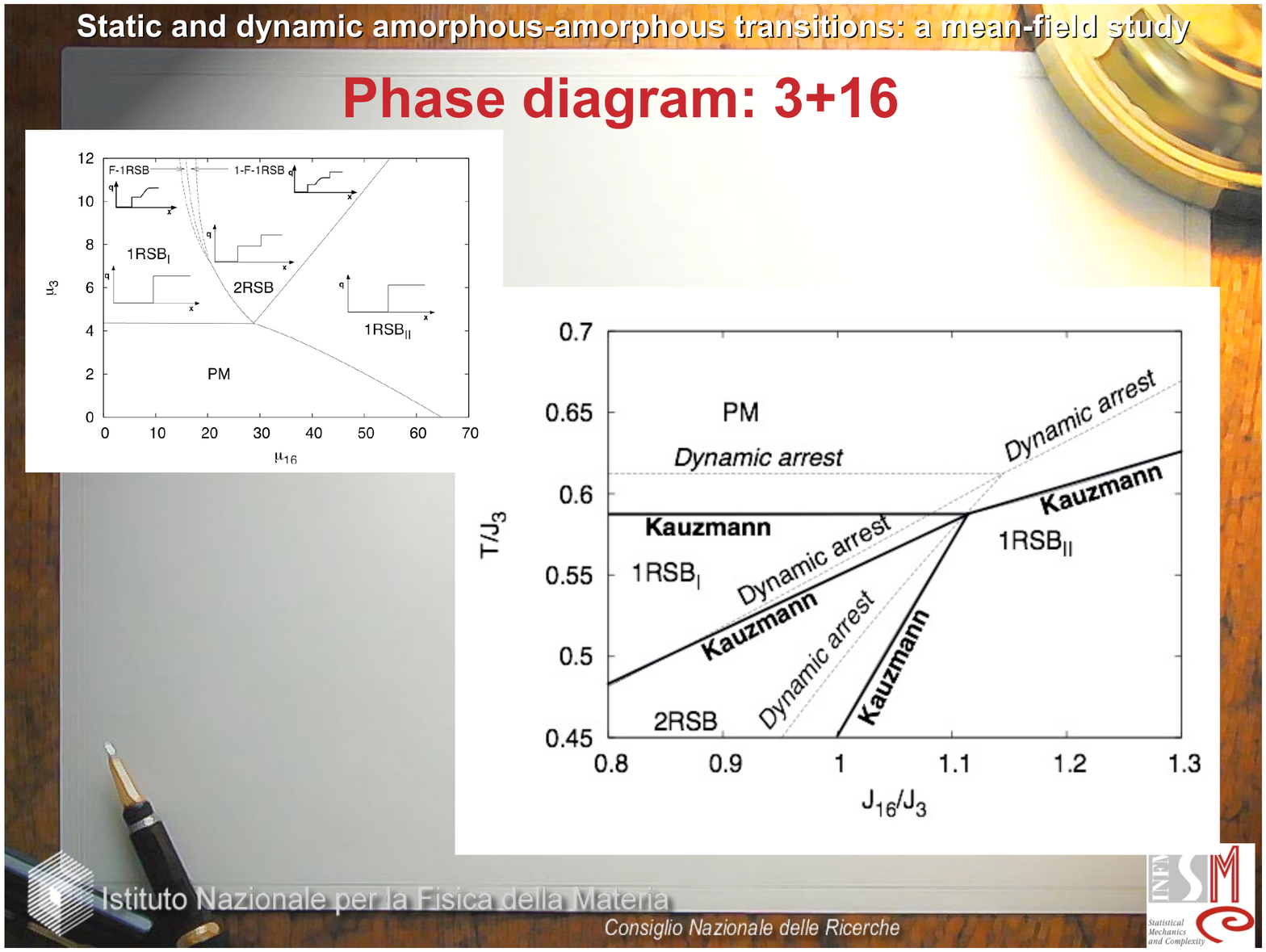}
\end{center}
\vskip  0 mm
 \caption{Phase diagram of the $3+16$ model with both dymanic (dotted)
 and thermodynamic (full) transition lines. Four phases are
present: paramagnetic (PM) at high $T$ and three glass phases, here termed
1RSB$_I$, 1RSB$_{II}$ and 2RSB (see text).}
\label{fig:phdi}
 \end{minipage}
\hspace*{.02\textwidth}
\begin{minipage}[b]{.44\textwidth}
\begin{center}
  \includegraphics[width=\textwidth]{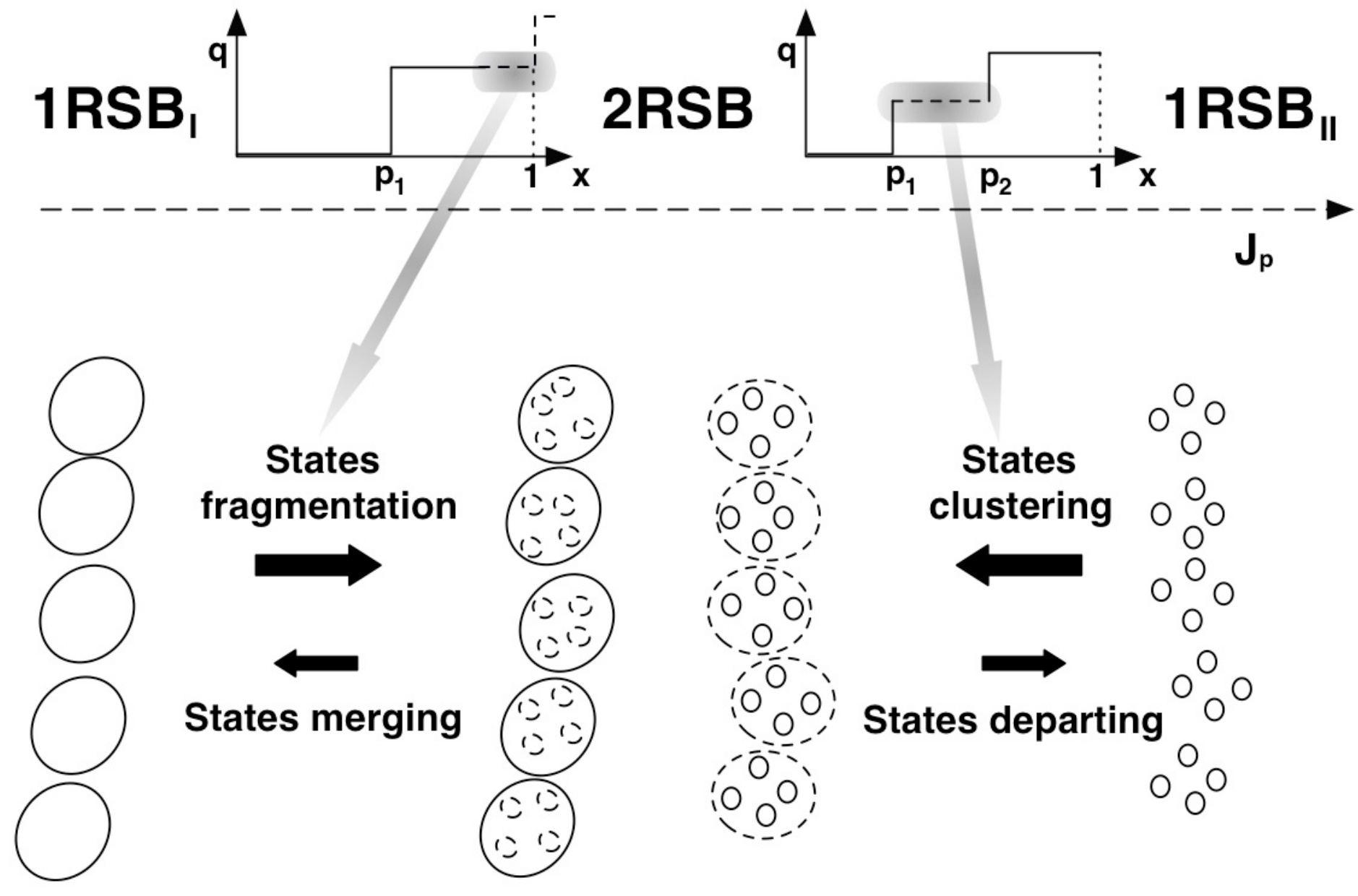}
\end{center}
\vskip  -3mm
 \caption{Interpretation of the GGT's in terms of metastable states
organization. In one case a GGT occurs when fast processes slow down
(and become secondary) and even faster processes appear (left,
``states fragmentation''); in the other case secondary processes show
up as new, intermediate, processes, between $\beta_{\rm fast}$ and
$\alpha$ processes.}
\label{fig:frag}
\end{minipage} 
\end{figure}

 In Fig. \ref{fig:phdi} we show a detail of the $(T/J_s,J_p/J_s)$
phase diagram of the $(s,p)=(3,16)$ model around the tricritical
point.  Both dynamic and thermodynamic (i.e., Kauzmann) transition
lines are plotted.\footnote{The thermodynamic transition, termed
``Kauzmann'' in the figure, is a so-called ``random first order
transition'', with no latent heat but a discontinuous order
parameter. This is an example of the mean-field scenario behind the
mosaic theory \cite{Lubchenko07}}
One can observe
that descreasing $T$ the dynamic transition between
glass phases of different nature always precedes the thermodynamic one.
The dynamic transition temperature $T_d(J_p)$ is the highest $T$ at
which the lifetime of high-lying local states becomes infinite and
their number grows like $\exp(S_c(T))$ with the size $N$.  The
Kauzmann temperature $T_K(J_p)$ is, instead, the highest $T$ at which
$S_c(f_{\rm eq})/N$ of the global {\em glassy}\footnote{At higher
temperature the global minima is a liquid/paramagnetic state.}  minima
goes to zero. The overlap order parameter jumps from $0$ to $q_{\rm
EA}$ while the free energy $\Phi$ is continuous: $\Phi_{\rm
liq}(T_K)=\Phi_{\rm glass}(T_K)$.

At low $T$, two glass to glass transitions (GGT) occur, at the lower
and higher $J_p$.  We notice thay they are not exactly of the same
kind, in terms of local states hierarchy change. We schematically
report in Fig.  \ref{fig:frag} how the states in the 1RSB phase
reorganize as the system transforms into a 2RSB glass in the two
cases.  In the transition at low $J_p$ a local state fragments into a
cluster of new local states, whereas accross the transition at high
$J_p$ subsets of uncorrelated local states group together in
correlated clusters.

In conclusion, we have examined a mean-field model displaying, in particular,
a phase whose thermodynamics is
described by a 2RSB solution ($q(x)=q_0,q_1,q_2$). 
Recalling both the thermodynamic and the dynamic properties of this specific 
phase (including the GGT's  from other glassy phases) and 
exploiting the  equivalence 
between RSB's and  time scale bifurcations, we argue that
\begin{enumerate}
\item changes connecting
two local states in the same cluster of states ($q=q_2$)
are  $\beta_{\rm fast}$ (else called $\gamma$) processes, 
\item changes connecting two local states in two different clusters
belonging to the same cluster of clusters ($q=q_1$) correspond to JG
processes,
\item changes
connecting  two uncorrelated states ($q=q_0\approx 0$) 
contribute to the $\alpha$
relaxation.
\end{enumerate}
The hierarchical nesting implicit in the present approach hints that
fast processes have a relevant influence on slow processes, even though 
taking place on well separated time-scales. This very
heuristic observation naturally stimulates a comparison with Ngai's
Coupling Model (see, e.g., Ref. \cite{Ngai} and references therein). A
study in this direction is in progress.

\end{document}